\documentclass[journal]{IEEEtran}

\ifCLASSINFOpdf
\else
   \usepackage[dvips]{graphicx}
\fi
\usepackage{url}

\usepackage{graphicx}
\usepackage{booktabs}
\usepackage{multirow}
\usepackage{threeparttable}

\begin{document}

\title{Exploring Speech Foundation Models for Speaker Diarization Across Lifespan}

\author{Anfeng Xu, \IEEEmembership{Graduate Student Member, IEEE}, Tiantian Feng \IEEEmembership{Member, IEEE}, \\Shrikanth Narayanan, \IEEEmembership{Fellow, IEEE}

\thanks{Anfeng Xu, Tiantian Feng, and Shrikanth Narayanan are all with University of Southern California (e-mail: anfengxu@usc.edu, tiantiaf@usc.edu, shri@usc.edu).}

\thanks{$\dagger$ Code and model weights available on https://github.com/usc-sail/Diarization-lifespan.}
}

\markboth{Journal of \LaTeX\ Class Files, Vol. 14, No. 8, August 2015}
{Shell \MakeLowercase{\textit{et al.}}: Bare Demo of IEEEtran.cls for IEEE Journals}
\maketitle

\begin{abstract}
Speech foundation models have shown strong transferability across a wide range of speech applications. However, their robustness to age-related domain shift in speaker diarization remains underexplored. In this work, we present a cross-lifespan evaluation within a unified end-to-end neural diarization framework (EEND-VC), covering speech samples from conversations involving children, adults, and older adults. We compare models under zero-shot cross-age inference, joint multi-age training, and domain-specific adaptation. Results show substantial performance degradation when models trained on adult-specific speech are applied to child and older-adult conversational data. Moreover, joint multi-age training across different age groups improves robustness without reducing diarization performance in canonical adult conversations, while targeted age group adaptation yields further gains in diarization performance, particularly when using the Whisper encoder.
\end{abstract}


\IEEEpeerreviewmaketitle

\section{Introduction}
Speaker diarization aims to automatically determine ``who spoke when and for how long?'' in multi-speaker recordings and serves as a fundamental component for downstream speech technologies, including automatic speech recognition (ASR)~\cite{park2022review}. The recent emergence of large-scale speech foundation models has reshaped the landscape of speech processing. Models such as Whisper~\cite{radford2023robust} and WavLM~\cite{chen2022wavlm}, trained on a massive amount of diverse speech data, have demonstrated strong transferability across a wide range of tasks beyond their original training objectives. However, their effectiveness for speaker diarization under substantial domain shifts, particularly across age groups, remains underexplored.

Most existing diarization benchmarks and systems are developed and evaluated primarily on ``standard" adult-specific speech corpora, typically from the broad middle life-span age groups between 25-60 years. While significant progress in speech technology has been achieved for these demographics, real-world applications often involve speakers that include children and older adults. Age-related developmental variability introduces substantial acoustic and conversational differences, such as changes in pitch range, articulation patterns, and speaking speed~\cite{lee1999acoustics, potamianos2004robust, smith1987temporal}. Likewise, older-adult speech exhibits variability in fluency and prosody, including altered rhythm, reduced pitch range and modulation, slower or fluctuating speech rate, and frequent word-finding pauses, which may reflect cognitive decline \cite{vigo2022speech}. As a result, diarization systems trained predominantly on the middle-band adult speech may perform poorly when applied to out-of-domain age groups. Systematic evaluation across age groups is essential for understanding model robustness and generalization.

End-to-end neural diarization (EEND)~\cite{fujita2019end, fujita2019endp}, especially when combined with vector clustering (EEND-VC)~\cite{kinoshita2021integrating, kinoshita2021advances}, has recently achieved competitive results on adult speech datasets. In parallel, speech foundation models such as Whisper and WavLM provide rich acoustic and linguistic representations that can be integrated into diarization architectures. While these models have demonstrated strong transferability across speech recognition and understanding tasks, their behavior under age-related domain shift has not been systematically compared within a unified diarization framework.

In this work, we conduct a comprehensive study of speech foundation models for speaker diarization involving speakers representing age groups from across the lifespan. Building on an EEND-VC framework, we leverage speech foundation models and evaluate them under three realistic scenarios: (1) zero-shot inference from adult-only training to child and older-adult speech, (2) joint multi-age training, and (3) domain adaptation via fine-tuning on target age groups. 

Our main contributions are summarized as follows:
\begin{itemize}
    \item We present a systematic cross-lifespan benchmark of speech foundation models for speaker diarization, covering adult, child–adult, and older-adult conversations$^\dagger$.

    \item We provide one of the first works of integrating Whisper encoders into an EEND-VC framework.

    \item We analyze zero-shot generalization, multi-domain training, and domain adaptation to understand how speech foundation models respond to age-related domain shifts.

    \item We show that adapting Whisper encoders benefits from domain adaptation, while adapting WavLM with strong diarization priors exhibits robust cross-age generalization.
\end{itemize}

\section{Method}
\subsection{EEND-VC Pipeline}
The proposed system is built upon DiariZen~\cite{han2025leveraging, han2025efficient}, which follows the EEND-VC framework. It uses the Pyannote~\cite{bredin2023pyannote} backend to cluster speakers and combine local EEND speaker diarization results. We vary the encoder of the EEND module across different speech foundation models, while keeping all other configurations identical.

The EEND module consists of an encoder, a Conformer~\cite{gulati2020conformer}, and a linear classification layer. Hidden representations from all encoder layers are aggregated using a learnable layer-wise weighted sum, which serves as input to the Conformer, consisting of 4 layers. Each Conformer layer includes a feed-forward module with input and hidden dimensions of 256 and 1024, respectively, a four-head multi-head self-attention module, and a convolution module with kernel size 31. All dropout rates are set to 0.1. The Conformer output is fed into a linear layer to generate frame-level logits, followed by a softmax that produces powerset labels for the EEND objective. The model is trained using the powerset loss, supporting up to four speakers with a maximum of two overlapping speakers.

For the vector clustering, we use agglomerative hierarchical clustering (AHC) for all experiments. Local speaker embeddings are extracted using a ResNet34LM model\footnote{Wespeaker/wespeaker-voxceleb-resnet34-LM from HuggingFace.} trained with the WeSpeaker toolkit~\cite{wang2024advancing} on the VoxCeleb2 dataset~\cite{chung2018voxceleb2}. Cosine similarity is used as the stopping criterion for cluster merging, with a similarity threshold of 0.70. In addition, we enforce a minimum cluster size of 30 segments. During clustering, the number of speakers is constrained to 2-8. 

\subsection{Speech Foundation Models for EEND module}

For the encoder of EEND, we explore speech foundation models from the WavLM and Whisper model families. 

\textbf{Whisper} is a transformer-based encoder–decoder model trained with weak supervision on 680k hours of multilingual speech data for ASR and related tasks. Its encoder has shown strong domain adaptation capabilities on other tasks, such as speech emotion recognition and dialect classification \cite{feng2025vox, feng2025voxlect}. Particularly, it has achieved competitive performance in child-adult speaker role diarization \cite{xu2024exploring, xu2025data}. We evaluate the Whisper encoders from the base, small, and medium variants \textbf{(Whisper-Base, Whisper-Small, and Whisper-Medium)}.

\textbf{WavLM} is a self-supervised speech model that employs k-means clustering to discretize speech representations following HuBERT~\cite{hsu2021hubert}, pre-trained with masked prediction. WavLM achieves strong performance across multiple speech recognition and understanding benchmarks~\cite{yang2021superb}. In our experiments, we use the Base+ and Large models \textbf{(WavLM-Base+, WavLM-Large)}, both pre-trained on 94k hours of audio data. We also evaluate the WavLM model released with DiariZen \textbf{(WavLM-DiariZen)}\footnote{BUT-FIT/diarizen-wavlm-large-s80-md-v2 from HuggingFace.}, which was fine-tuned on 8 datasets consisting of 637.6 training hours of multi-speaker conversational data and subsequently pruned. This version has demonstrated state-of-the-art performance in many speaker diarization datasets.

\section{Experiments}
\subsection{Datasets}

We use three widely adopted public datasets for general adult speaker diarization, along with one public dataset involving children and one involving older adults. Unless otherwise specified, we use the official train, dev, and test split from each dataset. The dataset details are in Table~\ref{tab:dataset-stats}. 

\subsubsection{Diarization datasets involving adult population}
For datasets primarily involving the adult population, we use far-field single-channel recordings from the public datasets AMI~\cite{kraaij2005ami}, AISHELL-4~\cite{fu2021aishell}, and AliMeeting~\cite{yu2022m2met}, similar to the work in \cite{han2025leveraging}. Together, these datasets cover both English and Mandarin conversations with 2 to 7 speakers. In the absence of an official development split for AISHELL-4, we follow \cite{han2025leveraging} for train and development data partitioning. The final training set is constructed by combining the training portions of all three datasets, and their respective development splits are also merged for validation. 

\subsubsection{Diarization datasets involving child population}
We use the Playlogue dataset~\cite{kalanadhabhatta2024playlogue}, which contains over 33 hours of naturalistic, long-form child and adult interactions from the TalkBank~\cite{macwhinney2007talkbank} system, spanning three play-based corpora and one narrative corpus with typically developing preschool children (3-5 years old). The dataset includes word-aligned transcripts generated using NVIDIA NeMo forced alignment~\cite{kuchaiev2019nemo}. Similar to \cite{xu2026end}, words with predicted durations longer than 2 seconds, which are mostly misaligned, are removed from both the audio and transcripts. Consecutive words from the same speaker with gaps under 0.3 seconds are merged. Playlogue remains challenging due to real-world recording conditions and less-accurate annotations stemming from the difficulty of forced alignment in child speech.

\subsubsection{Diarization datasets involving older adult population}
SeniorTalk~\cite{chen2025seniortalk} is a Mandarin conversational dataset containing over 55 hours of topic-driven, spontaneous dialogues from older adults (75–85 years old) across 16 provinces in China. The dataset provides long-form, two-speaker (both older adults) conversations, with speaker diarization annotations by human annotators. Collected in real-world settings, the dataset reflects diverse vocal characteristics of the older adult population. 

\begin{table}[t]
  \centering
  \footnotesize
  \setlength{\tabcolsep}{4pt}
  \caption{Dataset statistics.}
  \vspace{-3mm}
  \label{tab:dataset-stats}
  \begin{tabular}{lccc}
    \toprule
    \multirow{2}{*}{\textbf{Dataset}} & \multirow{2}{*}{\textbf{Age Group}} & \textbf{Hours} & \textbf{Number of Files} \\

    & & \textbf{(Train/Dev/Test)} & \textbf{(Train/Dev/Test)} \\
    \midrule
    AMI        & adult        & 79.7/9.7/9.1     & 134/18/16 \\
    AISHELL4   & adult        & 97.2/10.3/12.7   & 173/18/20 \\
    AliMeeting & adult        & 111.4/2.2/10.8   & 209/8/20 \\
    SeniorTalk & older adult  & 44.2/5.6/5.7     & 90/10/10 \\
    Playlogue  & child/adult  & 16.5/5.2/6.9     & 97/27/34 \\
    \bottomrule
  \end{tabular}
\end{table}

\begin{table*}[t]
  \centering
  \footnotesize
  \caption{DER (\%) on adult-only in-domain datasets and age-diverse (older adult and child) out-of-domain datasets. The best and second-best results are highlighted in bold and underline, respectively. }
  \vspace{-3mm}
  \label{tab:adult_results}
  \begin{tabular}{lccccccc}
     \toprule
     & & \multicolumn{4}{c}{\textbf{In-domain}} 
     & \multicolumn{2}{c}{\textbf{Out-of-domain}} \\
     & & \multicolumn{4}{c}{\textbf{Adult}} 
     & \textbf{Older Adult} & \textbf{Child/Adult} \\
     \cmidrule(lr){3-6} \cmidrule(lr){7-8}
    \textbf{EEND Encoder} & \textbf{Window} & AMI & AliMeeting & AISHELL4 & Macro Avg. & SeniorTalk & Playlogue \\
    \cmidrule(lr){1-2}\cmidrule(lr){3-6} \cmidrule(lr){7-8}
    WavLM-Base+   & 8s & 18.6 & 20.2 & 12.2 & 17.0 & 24.4 & 65.2 \\
    WavLM-Large  & 8s & 17.7 & 21.2 & 11.6 & 16.8 & 22.7 & 70.7 \\
    \cmidrule(lr){1-2}
    Whisper-Base     & 8s & 18.0 & 19.4 & 10.7 & 16.1 & 22.5 & 67.7 \\
    Whisper-Small    & 8s & 17.3 & 19.0 & 10.3 & 15.5 & 23.4 & 67.0 \\
    Whisper-Medium & 8s & \underline{16.2} & 18.0 & \textbf{9.8} & 14.7 & 22.1 & 72.0 \\
    Whisper-Medium & 16s & 17.0 & \underline{16.7} & \underline{10.1} & \underline{14.6} & \underline{21.4} & \underline{59.7} \\
    \cmidrule(lr){1-2}
    WavLM-DiariZen\textsuperscript{†}  & 16s & \textbf{13.7} & \textbf{12.1} & 10.4 & \textbf{12.0} & \textbf{18.0} & \textbf{53.2} \\
    \bottomrule
    \end{tabular}
    \begin{tablenotes}
    \footnotesize
    \centering
    \item \textsuperscript{†} Using public EEND module trained on a large compound dataset including the adult datasets above, without further training. AHC for Clustering.
    \end{tablenotes}

\end{table*}
\begin{table*}[t]
  \centering
  \footnotesize
  \caption{DER (\%) on adult-only and age-diverse (older adult and child) datasets under in-domain evaluation. The best and second-best results are highlighted in bold and underline, respectively. Values in parentheses () indicate the relative change (\%) compared to the adult-only training results from Table~\ref{tab:adult_results}.}
  \vspace{-3mm}
  \label{tab:in_domain_results}
    \begin{tabular}{lcccccccc}
    \toprule
     & & \multicolumn{6}{c}{\textbf{In-Domain}} \\
     \cmidrule(lr){3-8}
     & & \multicolumn{4}{c}{\textbf{Adult}} 
     & \textbf{Older Adult} & \textbf{Child/Adult} \\
    \cmidrule(lr){3-6} \cmidrule(lr){7-8}
    \textbf{EEND Encoder} & \textbf{Window} & AMI & AliMeeting & AISHELL4 & Macro Avg. & SeniorTalk & Playlogue \\
    \cmidrule(lr){1-2} \cmidrule(lr){3-6} \cmidrule(lr){7-8}
    WavLM-Base+  & 8s & 18.4 & 20.6 & 11.9 & 17.0 (+0.0\%) & 13.9 (-43.0\%) & 45.7 (-29.9\%) \\
    WavLM-Large & 8s & 17.9 & 20.0 & 11.4 & 16.4 (-2.3\%) & 13.4 (-41.0\%) & 46.5 (-31.1\%) \\
    \cmidrule(lr){1-2} 
    Whisper-Base    & 8s & 18.5 & 19.1 & 11.1 & 16.2 (+0.6\%) & 13.6 (-39.6\%) & 46.3 (-31.6\%) \\
    Whisper-Small   & 8s & 17.7 & 18.5 & \textbf{9.8} & 15.3 (-1.3\%) & 13.1 (-44.3\%) & 44.7 (-33.3\%)\\
    Whisper-Medium  & 8s & 16.7 & 17.6 & 10.1 & 14.8 (+0.7\%)  & 13.0 (-41.2\%) & 44.4 (-38.3\%) \\
    Whisper-Medium  & 16s & 16.5 & 16.1 & \textbf{9.8} & 14.1 (-3.2\%) & 11.6 (-45.8\%) & \underline{40.9 (-31.5\%)} & \\
    \cmidrule(lr){1-2}
    WavLM-DiariZen\textsuperscript{†} & 8s & \underline{14.5} & \underline{13.4} & 10.8 & \underline{12.9 (N/A)} & \textbf{11.4 (N/A)} & 43.5 (N/A) &  \\
    WavLM-DiariZen\textsuperscript{†} & 16s & \textbf{13.8} & \textbf{12.1} & 10.7 & \textbf{12.2 (+1.7\%)} & \textbf{11.4 (-36.7\%)} & \textbf{40.0 (-24.8\%)} \\
    \bottomrule
    \end{tabular}
    \begin{tablenotes}
    \footnotesize
    \centering
    \item \textsuperscript{†} Using public EEND module trained on a large compound dataset, further fine-tuned on the five listed datasets. AHC for Clustering.
    \end{tablenotes}

\end{table*}
\subsection{Evaluation Protocols}
\subsubsection{Adult-only training}
\label{subsec:zero-shot}
In the adult-only training setting, models are trained exclusively on the adult diarization datasets and evaluated without any further fine-tuning on the datasets involving child or older adult speech. Performance is first measured on the adult test sets and then evaluated directly on the child and senior datasets as out-of-domain test sets. This setting assesses the models' cross-age generalization to age-related acoustic and conversational domain shifts. For WavLM-DiariZen, we do not further train the model, as it is already pretrained on diverse adult diarization corpora, serving as a stronger adult-only baseline.

\subsubsection{Multi-age combined training}
In the multi-age combined training setting, we train on the combined adult, child, and older-adult diarization datasets and evaluate on the respective test splits of each population. This setting reflects scenarios where labeled data from children and older adults are available during training, enabling in-domain evaluation for these age groups. In addition, it allows us to analyze whether including child and older-adult speech affects performance on general adult conversations. 

\subsubsection{Domain adaptation by Age Group}
In the domain adaptation setting, we first train the model on the adult diarization datasets and then perform domain-specific fine-tuning on the child and older-adult datasets separately. That is, we start with the adult-trained model and adapt it to each target age group using its respective training split. Evaluation is conducted on the corresponding child and older-adult test sets. This setting assesses the effectiveness of age-specific adaptation relative to a general adult model and quantifies the gains from targeted fine-tuning under domain shift.

\subsection{Experimental Details}
For Sections~\ref {sec:zero-shot}, \ref{sec:in-domain}, and \ref{sec:domain-adapt}, we only fine-tune the Conformer and linear classification layers while the speech foundation model encoders are frozen. We train using the AdamW optimizer with a learning rate of $1e^{-3}$ and a batch size of 16 for 30 epochs. We validate the model at the end of each epoch and select the model with the lowest validation loss. We use the same setup for finetuning the encoder with LoRA \cite{hu2022lora} as in Section ~\ref{sec:finetune}, applying LoRA with rank 16 to the feed-forward layers of the transformer. Training segments are 8s or 16s long, with hop sizes of 6s and 12s, respectively. During inference, the same window length is used, with a hop size equal to the window size. For fine-tuning with the encoder unfrozen, as in Section ~\ref{sec:finetune}, we use a learning rate of $2e^{-5}$ for the encoder and $1e^{-3}$ for the Conformer and linear classification layers, while all other settings remain the same. We report Diarization Error Rate (DER) as the evaluation metric, with a 0s forgiveness collar and overlap included. We use a single NVIDIA A40 GPU for all the experiments.

\section{Results and Analysis}
\subsection{Results from Adult-only Training}
\label{sec:zero-shot}
Table~\ref{tab:adult_results} reports the DERs when the models are trained only on adult datasets. Among the WavLM and Whisper models trained from scratch for speaker diarization, Whisper-Medium achieves the best overall performance across the adult in-domain datasets. Out-of-domain results show heterogeneous trends, suggesting that models trained exclusively on adult speech exhibit limited generalization to age-diverse datasets. 

Overall, WavLM-DiariZen yields the strongest cross-age performance, likely due to large-scale training on speaker diarizatin datasets encompassing a broader speaker population. Although its training data primarily consists of adult speech, it includes a small portion of child–adult conversational data from DIHARD3~\cite{ryant2020third}, which may partially improve the performance on Playlogue. Nevertheless, although WavLM-Diarizen generalizes well to out-of-domain adult corpora~\cite{han2025efficient}, substantial errors persist in out-of-domain age groups, highlighting a continuing domain shift across age distributions.

\subsection{Results from Multi-Age Combined Training}
\label{sec:in-domain}

Table~\ref{tab:in_domain_results} presents results after training on all five datasets. Across all model variants, we observe substantially reduced DERs on age-diverse datasets, particularly in older adult and child–adult speech. Importantly, these gains on age-diverse speech are achieved while maintaining competitive performance on the general adult datasets. 

\subsection{Results from Domain Adaptation}
\label{sec:domain-adapt}

\begin{figure}[t]
  \centering
  \centerline{\includegraphics[width=0.75\linewidth]{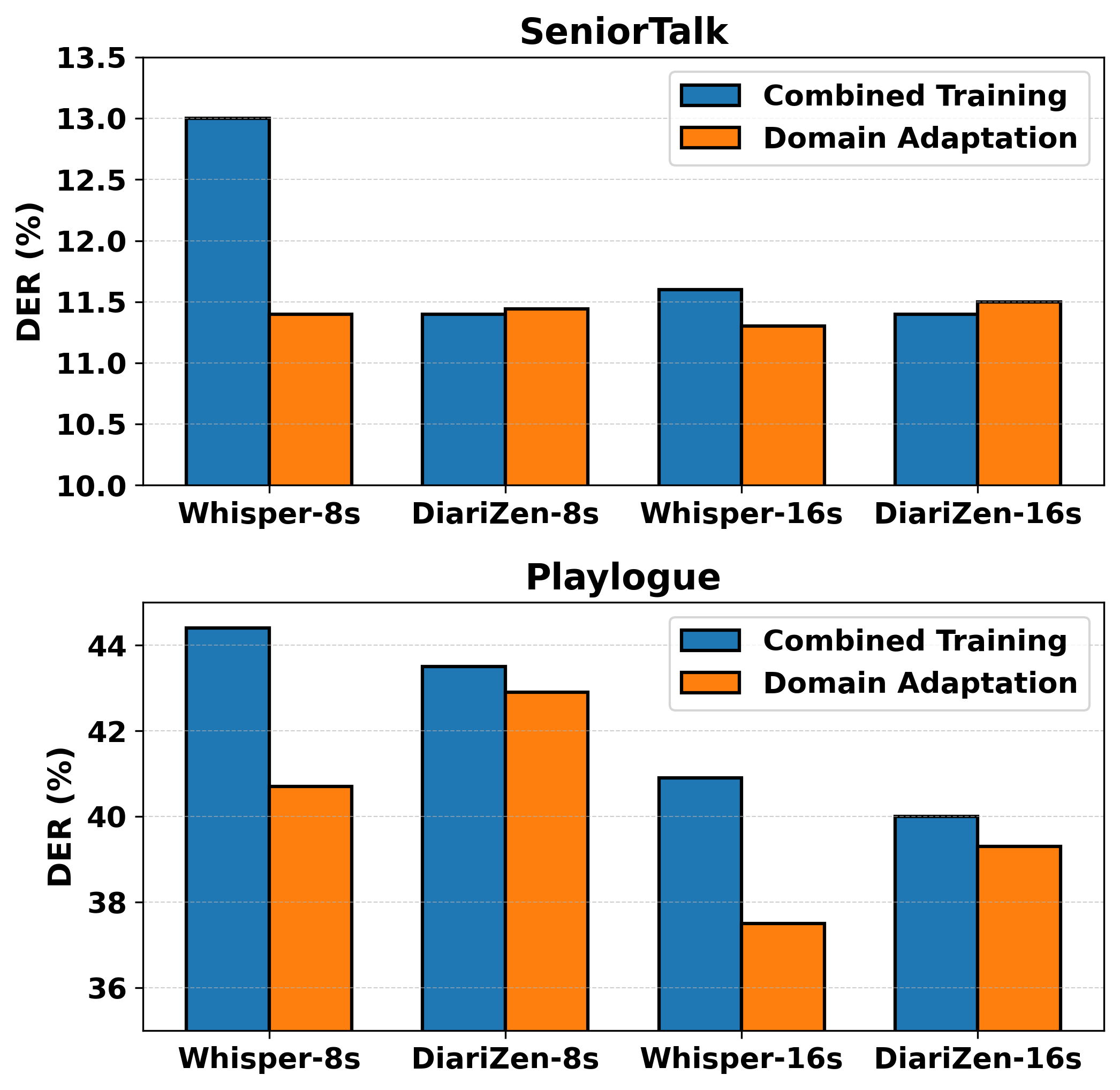}}
  \vspace{-3mm}
  \caption{DER (\%) under combined training and domain adaptation using Whisper-Medium and WavLM-DiariZen under 8s and 16s windows.}
  \label{fig:domain_adapt}
  \vspace{-2mm}
\end{figure}

Figure~\ref{fig:domain_adapt} shows DER after domain adaptation on SeniorTalk and Playlogue, compared to the multi-age joint training results in Table~\ref{tab:in_domain_results}. Domain adaptation yields clear additional gains for Whisper-Medium. Under both 8s and 16s windows, it consistently outperforms WavLM-DiariZen, with especially large improvements on Playlogue. With a 16s window, Whisper-Medium achieves the lowest DER on both datasets. This indicates that explicitly adapting to the target age distribution is more effective than relying on joint training across heterogeneous datasets. In contrast, WavLM-Diarizen shows limited performance gains compared to the in-domain fine-tuning.

We reason that this is because Whisper is pre-trained on larger-scale and diverse speech corpora, providing rich acoustic and linguistic representations that can be effectively specialized for target age groups.  In contrast, although WavLM-DiariZen benefits from extensive speaker diarization training, its diarization components are primarily optimized on adult corpora, and the underlying WavLM representations are less exposed to age-diverse speech. This may limit its flexibility when adapting to substantially different age distributions.



\subsection{Error Analysis under Cross-Age Domain Shift}

Table~\ref{tab:error_analysis} presents the decomposition of DER into missed detection (MD), false alarm (FA), and speaker confusion (SC) rates under adult-only training and domain adaptation settings. High MD rates are mainly due to cross-age acoustic differences, as child and older adult speech differ substantially from adult speech, making speech segments harder to detect under adult-trained models. In contrast, high FA errors are likely from noisier settings. Across both datasets, domain adaptation substantially reduces MD and FA errors, reflecting improved speech activity detection, particularly for child and older-adult speech. While SC rates increase slightly after adaptation, this is likely due to the reduced MD.

\begin{table}[t]
  \centering
  \footnotesize
  \caption{DER decomposition (\%) for SeniorTalk and Playlogue under adult-only training and domain adaptation settings. Whisper-Medium encoder with 8s window is used.}
  \vspace{-3mm}
  \label{tab:error_analysis}
    \begin{tabular}{llcccc}
    \toprule
    \textbf{Dataset} & \textbf{Setting} & \textbf{MD} & \textbf{FA} & \textbf{SC} & \textbf{DER} \\
    \midrule
    \multirow{2}{*}{SeniorTalk}
    & Adult-only   & 5.8  & 11.6 & 4.7 & 22.1    \\
    & Domain-Adapt & 1.0  & 2.8  & 7.4 & 11.2 \\
    \midrule
    \multirow{2}{*}{Playlogue}
    & Adult-only   & 26.8 & 37.6 & 7.7 & 72.0    \\
    & Domain-Adapt & 15.0 & 15.9 & 9.8 & 40.7 \\
    \bottomrule
    \end{tabular}
  \end{table}
\label{sec:finetune}

\subsection{Analysis on Whisper Encoder Fine-tuning}

\begin{table}[!t]
  \centering
  \footnotesize
  \setlength{\tabcolsep}{3pt}
  \caption{DER (\%) using Whisper-Medium encoder and 8s window under different fine-tuning (FT) strategies. Values in parentheses () indicate relative change from Whisper-Medium encoder frozen results under the same training settings.}
  \vspace{-3mm}
  \label{tab:training_comparison}
  \begin{tabular}{llccc}
    \toprule
    & & \textbf{Adult} & \textbf{Older Adult} & \textbf{Child/Adult} \\
    \textbf{Train} & \textbf{FT} & \textbf{Macro Avg. } & \textbf{SeniorTalk} & \textbf{Playlogue} \\
    \midrule
    Adult-only & LoRA     & 14.8(+0.7\%) & 22.2(+4.5\%) & 67.6(-6.2\%) \\
    & Updated & 15.5 (+5.4\%) & 22.2 (+0.5\%) & 67.6 (+6.1\%) \\
    \midrule
    Combined   & LoRA     & 14.6(-1.4\%) & 11.6(-10.8\%) & 41.9 (-5.6\%) \\
    & Updated & 15.1(+2.0\%)  & 12.0(-7.7\%)  & 45.8(+3.2\%)  \\
    \midrule
    Domain-Adapt      & LoRA     & N/A   & 11.3(+0.9\%)  & 40.0(-1.7\%)  \\
    & Updated & N/A   & 11.4(+1.8\%)  & 41.6(+2.2\%) \\
    \bottomrule
  \end{tabular}
\end{table}

Table~\ref{tab:training_comparison} compares LoRA \cite{hu2022lora} and full-parameter (Updated) fine-tuning using the Whisper-Medium encoder, which outperforms WavLM counterparts under the same training setups. Relative changes are reported with respect to the respective frozen-encoder baselines. LoRA produces little improvement in adult-only and domain-adaptation settings, but yields clearer gains in combined training, particularly on older adult and child–adult datasets. This suggests that lightweight adaptation is sufficient to specialize Whisper’s representations when supervision spans multiple age groups.

In contrast, full-parameter updating generally increases the errors. Given Whisper’s large-scale pre-training, extensive parameter updates may disrupt its prior representations, especially under limited or age-imbalanced supervision. Overall, these results indicate that parameter-efficient adaptation through LoRA is more stable and better aligned with cross-age generalization than full encoder updating.

\bibliographystyle{IEEEtran}
\bibliography{mybib}

\end{document}